\documentclass[a4paper,fleqn,usenatbib%,letters
]{mnras}
\pdfoutput=1

\usepackage[pdftex]{graphicx,color}
\usepackage[T1]{fontenc}
\usepackage[utf8]{inputenc}
\usepackage{lmodern}
\usepackage[normalem]{ulem}

\definecolor{urlblue}{rgb}{0,0,0.9}
\definecolor{linkgreen}{rgb}{0,0.45,0}
\definecolor{linkorange}{rgb}{0.7,0.1,0.0}

\usepackage{amsmath, amssymb}
\usepackage{cuted}
\usepackage[english]{babel}
\usepackage{enumerate}
\usepackage[normalem]{ulem}
\usepackage{enumitem}
\usepackage{multirow}
\usepackage{booktabs}
\usepackage{makecell}
\usepackage{bbold}

\setlist[enumerate]{wide=0pt, widest=99,leftmargin=\parindent, labelsep=* } 

\definecolor{valecol}{rgb}{0,0.5, 1.}

\definecolor{davidcol}{rgb}{0,0, 0.75}

\graphicspath{{./figs/}}

\AtBeginDocument{\hypersetup{
linkcolor=linkgreen,
citecolor=linkorange,
urlcolor=urlblue}}

\def\d{{\rm d}}

\title[On the use of the prior on the absolute magnitude of supernovae Ia]{On the use of the local prior on the absolute magnitude of Type Ia supernovae in cosmological inference}

\author[Camarena and Marra]{
David Camarena$^{1}$
and Valerio Marra$^{2,3,4}$
\\
$^{1}$PPGCosmo, Universidade Federal do Espírito Santo, 29075-910, Vitória, ES, Brazil\\
$^{2}$Núcleo de Astrofísica e Cosmologia \& Departamento de Física, Universidade Federal do Espírito Santo, 29075-910, Vitória, ES, Brazil\\
$^{3}$INAF -- Osservatorio Astronomico di Trieste, via Tiepolo 11, 34131, Trieste, Italy\\
$^{4}$IFPU -- Institute for Fundamental Physics of the Universe, via Beirut 2, 34151, Trieste, Italy
}

\date{Accepted XXX. Received YYY; in original form ZZZ}

\pubyear{202X}

\begin{document}
\label{firstpage}
\pagerange{\pageref{firstpage}--\pageref{lastpage}}

\maketitle

\begin{abstract}
A dark-energy which behaves as the cosmological constant until a sudden phantom transition at very-low redshift ($z<0.1$) seems to solve the >4$\sigma$ disagreement between the local and high-redshift determinations of the Hubble constant, while maintaining the phenomenological success of the $\Lambda$CDM model with respect to the other observables.
Here, we show that such a hockey-stick dark energy cannot solve the $H_0$ crisis.
The basic reason is that the supernova absolute magnitude $M_B$ that is used to derive the local $H_0$ constraint is not compatible with the $M_B$ that is necessary to fit supernova, BAO and CMB data, and this disagreement is not solved by a sudden phantom transition at very-low redshift.
We make use of this example to show why it is preferable to adopt in the statistical analyses the prior on $M_B$ as an alternative to the prior on $H_0$.
The three reasons are: i) one avoids potential double counting of low-redshift supernovae, ii) one avoids assuming the validity of cosmography, in particular fixing the deceleration parameter to the standard model value $q_0=-0.55$, iii) one includes in the analysis the fact that $M_B$ is constrained by local calibration, an information which would otherwise be neglected in the analysis, biasing both model selection and parameter constraints.
We provide the priors on $M_B$ relative to the recent Pantheon and DES-SN3YR supernova catalogs.
We also provide a Gaussian joint prior on $H_0$ and $q_0$ that generalizes the prior on $H_0$ by SH0ES.
\end{abstract}


%%%%%%%%%%%%%%%%%%%%%%%%%%%%%%%%%%%%
%%%%%%%%%%%%%%%%%%%%%%%%%%%%%%%%%%%%
\section{Introduction}\label{sec:intro}
%%%%%%%%%%%%%%%%%%%%%%%%%%%%%%%%%%%%
%%%%%%%%%%%%%%%%%%%%%%%%%%%%%%%%%%%%

The Hubble constant $H_0$ -- the first cosmographic coefficient in a series expansion of the scale factor -- is perhaps the most basic parameter in cosmology.
It is then understandable that the >4$\sigma$ disagreement between the local \citep{Riess:2020fzl} and high-redshift \citep{Aghanim:2018eyx} determinations of the Hubble constant has received much spotlight.
Indeed, this tension could very well signal the need of a new standard model of cosmology, although it is not clear which alternative model can  successfully explain all available observations~\citep[see][for details]{Knox:2019rjx,DiValentino:2021izs}.

A dark-energy which behaves as the cosmological constant until a sudden phantom transition at very-low redshift seems able to solve the $H_0$ crisis, while maintaining the phenomenological success of the $\Lambda$ cold dark matter (CDM) model with respect to the other observables.
The phenomenology of a late-time transition in the Hubble rate has been first considered by \citet{Mortonson:2009qq}, and recently confronted with data by \citet{Benevento:2020fev,Dhawan:2020xmp,Efstathiou:2021ocp}, while a low-redshift transition on the dark energy equation of state has been proposed by \citet{Alestas:2020zol} \citep[see also][]{Keeley:2019esp}.

Here, we show that a hockey-stick%
\footnote{We remind the reader that a hockey stick trend is characterized by a sharp change after a relatively flat and quiet period.}
dark energy ($hs$CDM, see Figure~\ref{fig:hs}) cannot solve the $H_0$ crisis.
The basic reason is that the supernova absolute magnitude $M_B$ that is used to derive the local $H_0$ constraint is not compatible with the $M_B$ that is necessary to fit supernova, baryon acoustic oscillations (BAO) and cosmic microwave background (CMB) data, and this remains true even with a sudden phantom transition at very-low redshift.
Statistically, this becomes evident if one includes the supernova calibration prior on $M_B$ in the statistical analysis, which would otherwise support $hs$CDM.

We make use of this example to show in details why it is preferable to adopt the prior on $M_B$ rather than the prior on $H_0$ in the cosmological analyses that study the impact of local $H_0$ on the dark energy properties \citep[see, for instance, the analysis performed in Section 5 of][]{Riess:2016jrr}. We also provide the $M_B$ priors relative to the Pantheon and Dark Energy Survey Supernova Program (DES-SN3YR) catalogs, and a joint prior on $H_0$ and $q_0$ that generalizes the one on $H_0$ by the Supernova H0 for the Equation of State (SH0ES) collaboration.

This paper is organized as follows.
In Section~\ref{sec:hockey} we introduce hockey-stick dark energy, in Section~\ref{sec:prior} we discuss the prior on $M_B$, while in Section~\ref{sec:bayes} we present the statistical analysis.
The results are shown in Section~\ref{sec:results} and the conclusions drawn in Section~\ref{sec:conclu}.

%%%%%%%%%%%%%%%%%%%%%%%%%%%%%%%%%%%%
%%%%%%%%%%%%%%%%%%%%%%%%%%%%%%%%%%%%
\section{Hockey-stick dark energy}\label{sec:hockey}
%%%%%%%%%%%%%%%%%%%%%%%%%%%%%%%%%%%%
%%%%%%%%%%%%%%%%%%%%%%%%%%%%%%%%%%%%

%%%%%%%%%%%%%%%%%%%%%%%%%%%%%%%%%%%%%%%%%%%%%%%%%%%%%    
\begin{figure}
\centering % trim={<left> <lower> <right> <upper>}
\includegraphics[trim={0 0 0 0}, clip, width=\columnwidth]{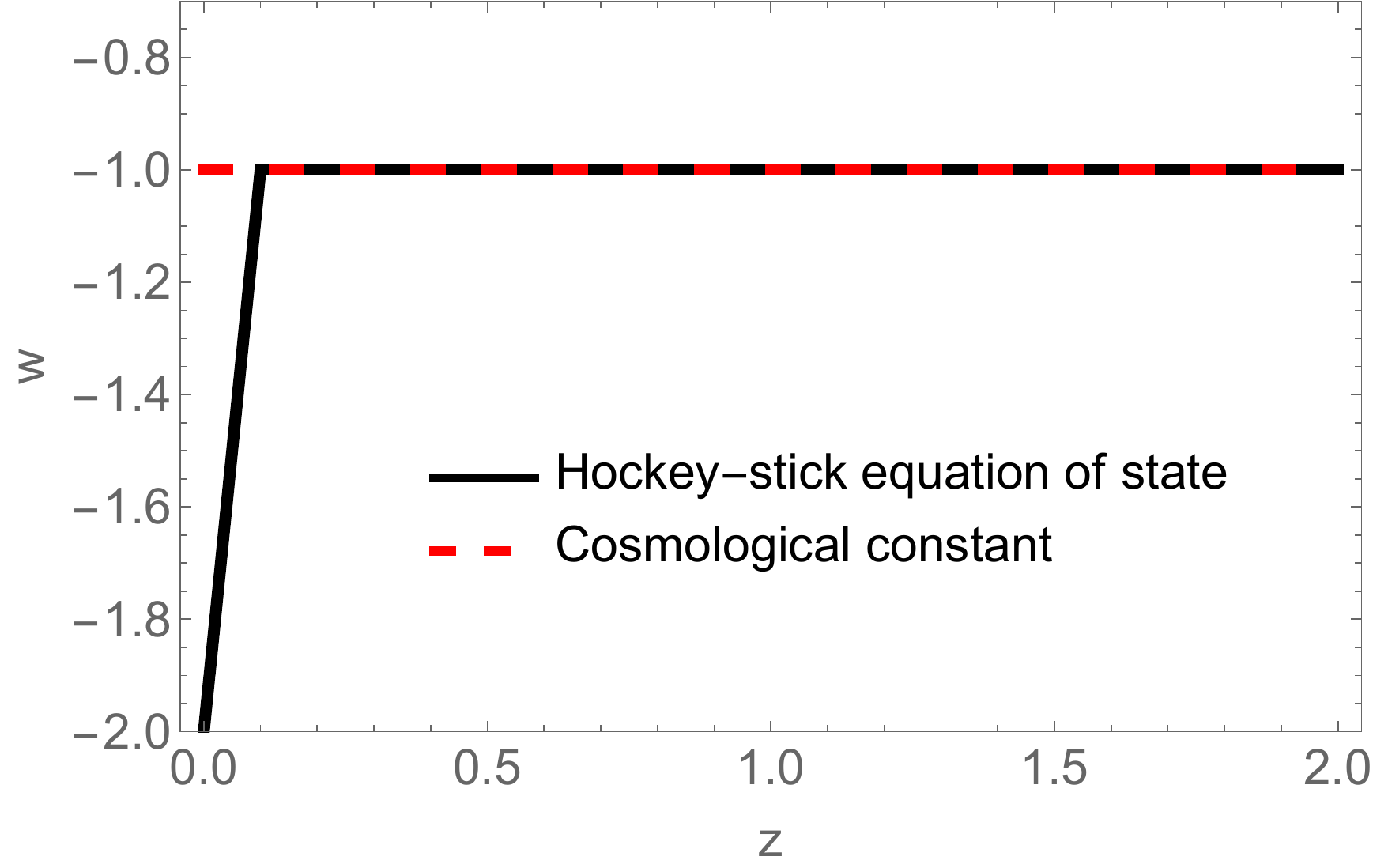}
\caption{Hockey-stick dark energy behaves as the cosmological constant until a sudden phantom transition at very-low redshift.}
\label{fig:hs}
\end{figure}
%%%%%%%%%%%%%%%%%%%%%%%%%%%%%%%%%%%%%%%%%%%%%%%%%%%%%

In order to show the advantages of using a local prior on $M_B$ instead of a local prior on $H_0$ we will consider a model that features a dark energy with the following hockey-stick equation of state ($hs$CDM):
\begin{equation}
w =\left\lbrace \!\!
\begin{array}{ll}
w_x-  (1+w_x)\, z/z_t & \text{ if } z\le z_t   \text{ (the blade)}\\
-1 & \text{ if }  z> z_t  \text{ (the shaft)}
\end{array}
\right. \!\!,
\label{eq:eos}
\end{equation} 
which mimics the cosmological constant at higher redshifts and deviates from the latter for $z\le z_t $, reaching $w_x$ at $z=0$, see Figure~\ref{fig:hs}.
A step equation of state (constant $w_x$ for $z\le z_t$) shows a very similar phenomenology. Here, we adopt the hockey-stick equation of state  as it features the same number of parameters ($w_x$ and $z_t$) but is continuous.
Models that feature the hockey-stick phenomenology are discussed in \citet{Mortonson:2009qq}.

It follows that the expansion rate is, assuming spatial flatness:
\begin{align}\label{hhh}
{H^2(z) \over H_{0}^{2} } %\equiv E^{2}(z)
= \Omega_{M0}   (1+z)^{3}+ \Omega_{R0}   (1+z)^{4} +  \Omega_{\Lambda0}  (1+z)^{3 g(z)}  ,  
\end{align}
where $\Omega_{M0} + \Omega_{R0}+  \Omega_{\Lambda0} =1$ %, $\Omega_{R0}= 4.18 \, h^{-2} 10^{-5}$,
and
\begin{align}
g(z)&= {1 \over \ln (1+z)} \int_{0}^{z}{1+w(z') \over 1+z'}dz'  \\
&= \frac{1+w_x}{z_t \ln (1+z)}  \times
\left\lbrace \!\!
\begin{array}{ll}
(1+z_t )\ln (1+z)-z & \text{if } z\le z_t \\
(1+z_t )\ln (1+z_t)-z_t & \text{if }  z> z_t
\end{array}
\right. \!\!. \nonumber
\end{align}
%
%\david{CLASS notation:
%\begin{align}
%\tilde{g}(a) & = 3\int_a^{a_0} \frac{1+w(a)}{a} da \,, \\
%\tilde{g}(a) & = \frac{3(1+w_x)}{(1-a_t)}  \times
%\left\lbrace \!\!
%\begin{array}{ll}
%-\ln a- (1-a)a_t/a & \text{if } a_t \le z \le a_0  \\
%-\ln a_t- (1-a_t)  & \text{if }   0 < a < a_t \\
%\end{array}
%\right. \!\!. \nonumber
%\end{align}
%Note that $\tilde{g}(z) = 3 \ln(1+z) g(z)$ and for  $a->1 $ we have $\tilde{g}(z) = 0$}.
%
The apparent magnitude is then:
\begin{equation} \label{mB}
m_B(z) = 5\log_{10}\left[\frac{d_L(z)}{10 \text{pc}} \right]  + M_B \,, 
\end{equation}
where the luminosity distance is:
\begin{align} \label{dL}
d_L(z) = (1+z) \int_0^z \frac{c \, \d \bar z}{H(\bar z)}  \,.
\end{align}
Finally, the distance modulus is given by:
\begin{align}
\mu(z)= m_B(z) - M_B \,.
\end{align}

For $z_t \rightarrow \infty$ one recovers the $w$CDM model with $w=w_x$.
We will consider the $w$CDM model for comparison sake.

%%%%%%%%%%%%%%%%%%%%%%%%%%%%%%%%%%%%
%%%%%%%%%%%%%%%%%%%%%%%%%%%%%%%%%%%%
\section{Supernova calibration prior}\label{sec:prior}
%%%%%%%%%%%%%%%%%%%%%%%%%%%%%%%%%%%%
%%%%%%%%%%%%%%%%%%%%%%%%%%%%%%%%%%%%

The determination of $H_0$ by the SH0ES Collaboration is a two-step process \citep{Riess:2016jrr}:
\begin{enumerate}
\item First, anchors, Cepheids and calibrators are combined to produce a constraint on the supernova Ia absolute magnitude $M_B$. This step only depends  on the astrophysical properties of the sources.
\item Second, Hubble-flow Type Ia supernovae in the redshift range $0.023 \le z \le 0.15$ are used to probe the luminosity distance-redshift relation in order to determine $H_0$. Cosmography with $q_0=-0.55$ and $j_0=1$ is adopted.
\end{enumerate}
The latest constraint by SH0ES reads:
\begin{align} \label{R21}
H^{\rm R21}_0 = 73.2 \pm 1.3 \text{ km s}^{-1} {\rm Mpc}^{-1}  \text{ \citep{Riess:2020fzl}} \,.
\end{align}

Usually, one introduces in the cosmological  analyses that use an $H_0$ prior the following $\chi^2$ function:
\begin{align} \label{chi2H}
\chi^2_{H_0}  = \frac{\big (H_0-H^{\rm R21}_0 \big)^2}{\sigma_{H_0^{\rm R21}}^2} \,.
\end{align}
The goal of this paper is to show, using the example of hockey-stick dark energy, that it is preferable to skip  step ii) above and adopt directly the local prior on $M_B$ via:
\begin{align} \label{chi2M}
\chi^2_{M_B}  = \frac{\big (M_B-M_B^{\rm R21} \big )^2}{\sigma_{M_B^{\rm R21}}^2} \,,
\end{align}
where $M_B^{\rm R21}$ is the calibration that corresponds to the $H_0$ prior of equation~\eqref{R21}.

Before proceeding, it is important to point out that supernovae Ia become standard candles only after standardization and that the method used to fit supernova Ia light curves, and its parameters, can influence the inferred value of $M_B$ (e.g., $x_0$, $x_1$ and $c$ in the case of SALT2, \citealt{Guy:2007dv}).
This means that the actual prior on $M_B$ from SH0ES can only be used with the Supercal supernova sample~\citep{Scolnic:2015eyc}, which is the one adopted by SH0ES in the latest analyses.

Consequently, in order to meaningfully use the local prior on $M_B$, one has to translate it to the light curve calibration adopted by some other dataset X.
This task can be achieved using the method developed in \citet{Camarena:2019moy}: 
the basic idea is to demarginalize the final $H_0$ measurement using for step ii) the supernovae of the dataset X that are in the same redshift range $0.023 \le z \le 0.15$.
This procedure, applied to the latest supernova catalogs, produces the priors listed in Table~\ref{calib}.
In other words, by adopting the priors given in Table~\ref{calib} and performing step ii), one recovers the $H_0$ determination of equation~\eqref{R21}.

It is worth mentioning that supernovae Ia are not perfectly standardizable candles and there are residual correlations with their environment, such as the step correction to $M_B$ according to the host galaxy mass \citep{2010ApJ...715..743K,2010ApJ...722..566L,2010MNRAS.406..782S}. The method discussed in this section assumes that these residual corrections have been applied before obtaining the effective prior on~$M_B$.

Correlations between the residuals and the supernova environment have also been used to argue in favor of a possible time evolution of the absolute magnitude \citep{Kang:2019azh,Kim:2019npy}. Recent analyses suggest that such time evolution is not favored by data \citep{Huang:2020mub,Koo:2020ssl,Sapone:2020wwz} and could have been produced by systematics \citep{Brout:2020msh,Rose:2020shp}. Throughout this work we assume that  $M_B$ does not evolve with time.

%%%%%%%%%%%%%%%%%%%%%%%%%%%%%%%%%%%%%%%%%%%%%%%%%%%%%
\begin{table}
\begin{center}
\setlength{\tabcolsep}{5pt}
\renewcommand{\arraystretch}{1.5}
\begin{tabular}{lcc}
\hline
\hline
SN dataset & Reference & Effective prior on $M_B^{\rm R21}$ \\
\hline
Supercal & \citet{Scolnic:2015eyc} & $ -19.2421 \pm 0.0375$ mag  \\
Pantheon &  \citet{Scolnic:2017caz} & $ -19.2435 \pm 0.0373$ mag \\
DES-SN3YR & \citet{Brout:2018jch} & $-19.2389 \pm 0.0336$ mag \\
\hline
\hline
\end{tabular}
\caption{
Effective prior on $M_B$ to be used instead of the prior on $H_0$ by  \citet{Riess:2020fzl} when carrying out cosmological inference with the corresponding supernova dataset.
Code and up-to-date values available at \href{https://github.com/valerio-marra/CalPriorSNIa}{github.com/valerio-marra/CalPriorSNIa}.
}
\label{calib}
\end{center}
\end{table}
%%%%%%%%%%%%%%%%%%%%%%%%%%%%%%%%%%%%%%%%%%%%%%%%%%%%%

%%%%%%%%%%%%%%%%%%%%%%%%%%%%%%%%%%%%
\subsection{Local joint $H_0$-$q_0$ constraint}\label{sec:astro}
%%%%%%%%%%%%%%%%%%%%%%%%%%%%%%%%%%%%

Although, as we argue below, it is preferable to use in the statistical analysis the prior on $M_B$, it is nevertheless important to determine the local value of $H_0$.

The measurement by SH0ES of equation \eqref{R21} is obtained from the local constraint  on $M_B$ after adopting in the cosmographic analysis  the following Dirac delta prior on the deceleration parameter $q_0$ :
\begin{align}
f(q_0) &= \delta (q_0 -q_{0, \rm{fid}}) \,, \\
q_{0, \rm{fid}} &=\frac{3}{2}\Omega_{M0, \rm{fid}}-1 = -0.55 \,,
\end{align}
where the deceleration parameter takes the value relative to the flat concordance $\Lambda$CDM model with $\Omega_{M0, \rm{fid}}=0.3$ \citep{Riess:2016jrr}. In other words, the constraint of equation \eqref{R21} uses information beyond the local universe in order to fix the value of $q_0$.

\begin{figure}
\centering
\includegraphics[width=\columnwidth]{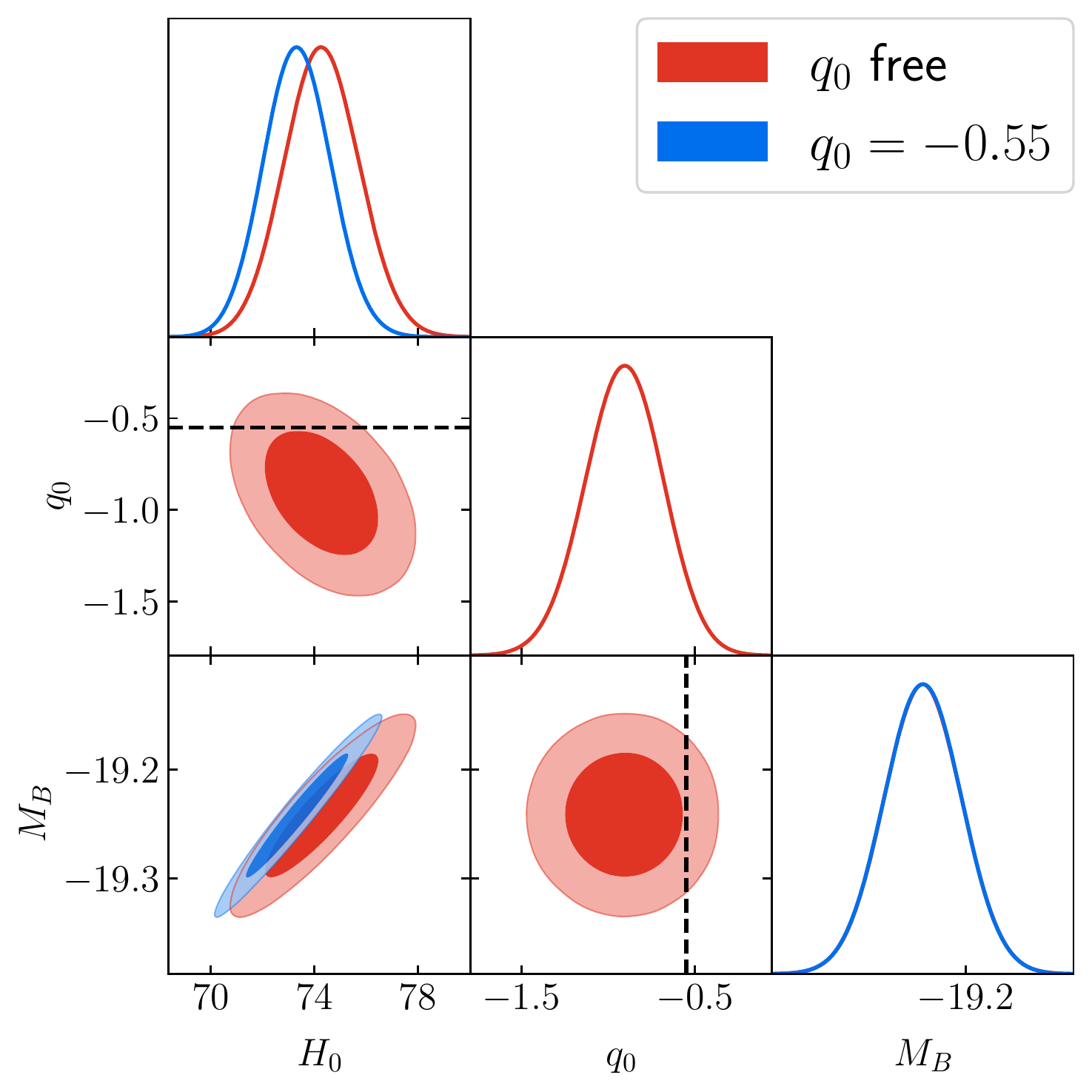}
\caption{Marginalized local constraints on $H_0$, $q_0$ and $M_B$ (in red). If the deceleration parameter is fixed to $q_0=-0.55$,  one obtains the original $H_0$ constraint by SH0ES (in blue).}
\label{triplot:cm}
\end{figure}

One can improve the local determination of $H_0$ by adopting an uninformative prior $f(q_0) = \text{constant}$.
Specifically, adopting  the $M_B$ prior relative to the Supercal dataset given in Table~\ref{calib} and the same 217 Supercal supernovae used by SH0ES, one obtains the joint prior that is given in Table~\ref{CM20} and illustrated in Figure~\ref{triplot:cm}.
This constraint on $H_0$ and the CMB-only constraint from the Planck Collaboration \citep{Aghanim:2018eyx} disagree at the 4.5$\sigma$ level.
We have used the numerical codes \texttt{emcee} \citep{ForemanMackey:2012ig} and \texttt{getdist} \citep{Lewis:2019xzd}.

It is worth noting that the determination of Table~\ref{CM20} only assumes large-scale homogeneity and isotropy and no information from observations beyond the local universe is used.
For comparison, we show in Figure~\ref{triplot:cm} also the  original constraint of equation \eqref{R21} that is recovered by fixing $q_0=-0.55$.%
\footnote{To be precise, the constraint of equation \eqref{R21} adopts third-order cosmography and fixes also $j_0=1$.
As in Figure~\ref{triplot:cm} we use second-order cosmography, 
fixing $q_0=-0.55$ gives back an $H_0$ that is $0.1 \text{ km s}^{-1} {\rm Mpc}^{-1}$ higher than the one of equation~\eqref{R21}.}
Note also that $M_B$ shows basically no correlation with~$q_0$. In other words, fixing $q_0=-0.55$ \citep{Riess:2020fzl}   should not have biased the determination of~$M_B$ via the method of \citet{Camarena:2019moy}.

%%%%%%%%%%%%%%%%%%%%%%%%%%%%%%%%%%%%%%%%%%%%%%%%%%%%%
\begin{table}
\begin{center}
\setlength{\tabcolsep}{10pt}
\renewcommand{\arraystretch}{1.5}
\begin{tabular}{lc|cc}
\hline
\hline
\multirow{1}{*}{parameter} & \multirow{1}{*}{$\mu_i \pm \sigma_i $} & \multicolumn{2}{c}{$C_{ij}$}  \\
%parameter & $\mu_i \pm \sigma_i $& $H_0$ & $q_0$ \\
\hline
$H_0 \;\;[\frac{\rm km/s}{\rm Mpc}]$ & $74.30 \pm 1.45$  & 1 & -0.41 \\
$q_0$ & $-0.91  \pm 0.22$ & -0.41 & 1 \\
\hline
\hline
\end{tabular}
\caption{
Joint prior on $H_0$ and $q_0$, marginalized over the supernova absolute magnitude $M_B$.
$\mu_i \pm \sigma_i $ are the marginalized mean values and standard deviation for the parameters, while $C_{ij}$ is the correlation matrix. The covariance matrix is given by $\Sigma_{ij} = \sigma_i \sigma_j C_{ij}$ (without summation).
This determination of $H_0$ generalizes the one of equation \eqref{R21} by SH0ES.
}
\label{CM20}
\end{center}
\end{table}
%%%%%%%%%%%%%%%%%%%%%%%%%%%%%%%%%%%%%%%%%%%%%%%%%%%%%

%%%%%%%%%%%%%%%%%%%%%%%%%%%%%%%%%%%%%%%%%%%%%%%%%%%%%    
%%%%%%%%%%%%%%%%%%%%%%%%%%%%%%%%%%%%%%%%%%%%%%%%%%%%%

%%%%%%%%%%%%%%%%%%%%%%%%%%%%%%%%%%%%
%%%%%%%%%%%%%%%%%%%%%%%%%%%%%%%%%%%%
\section{Statistical inference}\label{sec:bayes}
%%%%%%%%%%%%%%%%%%%%%%%%%%%%%%%%%%%%
%%%%%%%%%%%%%%%%%%%%%%%%%%%%%%%%%%%%

We now discuss the datasets that we adopt in order to constrain the $hs$CDM model.

%%%%%%%%%%%%%%%%%%%%%%%%%%%%%%%%%%%%
\subsection{Cosmic Microwave Background}
%%%%%%%%%%%%%%%%%%%%%%%%%%%%%%%%%%%%

We use the Gaussian prior on $(R, l_a,\Omega_{B0}\, h^2, n_s)$ derived from the Planck 2018 results \citep[][$w$CDM model in Table~I]{Chen:2018dbv}.
We denote with $\chi^2_{\rm cmb}$ the corresponding~$\chi^2$ function.

%%%%%%%%%%%%%%%%%%%%%%%%%%%%%%%%%%%%
\subsection{Baryonic Acoustic Oscillations}
%%%%%%%%%%%%%%%%%%%%%%%%%%%%%%%%%%%%

We  adopt BAO measurements from the following surveys: 6dFGS \citep{Beutler:2011hx}, SDSS-MGS \citep{Ross:2014qpa} and BOSS-DR12 \citep{Alam:2016hwk}. 6dFGS and SDSS-MGS provide isotropic measurements at redshifts $0.1$ and $0.15$, while BOSS-DR12 data constrains $H(z)$ and $d_A(z)$ at redshifts $0.38$, $0.51$ and $0.61$.
We denote with $\chi^2_{\rm bao}$ the corresponding~$\chi^2$ function.

%%%%%%%%%%%%%%%%%%%%%%%%%%%%%%%%%%%%
\subsection{Supernovae Ia}
%%%%%%%%%%%%%%%%%%%%%%%%%%%%%%%%%%%%

We consider the Pantheon dataset, consisting of 1048  Type Ia supernovae spanning the redshift range $0.01 <z< 2.3$~\citep{Scolnic:2017caz}.
We denote with $\chi^2_{\rm sne}$ the corresponding~$\chi^2$ function.

%%%%%%%%%%%%%%%%%%%%%%%%%%%%%%%%%%%%
\subsection{Local constraint}
%%%%%%%%%%%%%%%%%%%%%%%%%%%%%%%%%%%%

We will consider either the prior on $H_0$ of equation~\eqref{chi2H} or the prior on $M_B$ of equation~\eqref{chi2M} relative to the Pantheon sample, see Table~\ref{calib}.

%%%%%%%%%%%%%%%%%%%%%%%%%%%
%%%%%%%%%%%%%%%%%%%%%%%
\subsection{Total likelihood: \boldmath{$M_B$} vs \boldmath{$H_0$}}

The main goal of this paper is to show how the result of the analysis is biased when using $\chi^2_{H_0}$ instead of $\chi^2_{M_B}$. To this end we will build and compare the following two likelihoods:
\begin{align} 
\chi^2_{{\rm tot}, H_0} (\theta) %\equiv  -2 \ln \mathcal{L}_{\text{tot},H_0}
& =  \chi^2_{\rm cmb}   + \chi^2_{\rm bao} 
+ \chi^2_{\rm sne}+ \chi^2_{H_0}  \,,  \label{chi2totH} \\
\chi^2_{{\rm tot}, M_B} (\theta) %\equiv  -2 \ln \mathcal{L}_{\text{tot},M_B} 
&=  \chi^2_{\rm cmb}   + \chi^2_{\rm bao} 
+ \chi^2_{\rm sne}+ \chi^2_{M_B}   \,. \label{chi2totM}
\end{align}
Note that the number of data points is the same for both analyses.

In both cases the parameter vector is:
\begin{align}
\theta = \{H_0, \Omega_{M0}, w_x, z_t, M_B, \Omega_{B0}, n_s   \} \,.
\end{align}
In particular, the posteriors are not marginalized analytically over $M_B$ so that we can obtain the posterior on $M_B$.
However, it is often computationally useful to marginalize the posterior over $M_B$ and in the next Section we present the corresponding formulas.

%%%%%%%%%%%%%%%%%%%%%%%%%%%
%%%%%%%%%%%%%%%%%%%%%%%
\subsection{Posterior marginalized over $M_B$}

In the case of the $\chi^2_{{\rm tot}, H_0}$ of equation~\eqref{chi2totH}, it is well known that one can marginalize analytically the posterior over $M_B$ \citep{Goliath:2001af}. As we will show below, this is possible also for the $\chi^2_{{\rm tot}, M_B} $ of equation~\eqref{chi2totM}. For completeness we will present both cases.

%%%%%%%%%%%%%%%%%%%%%%%%%%%%%%%%%%%%
\subsubsection{Prior on $H_0$}
%%%%%%%%%%%%%%%%%%%%%%%%%%%%%%%%%%%%

Since, in this case, $M_B$ enters only the SN likelihood, we will consider only the latter.
The $\chi^{2}$ function is:
\begin{align}
\chi_{\text{sne}}^{2}&= \{ m_{B,i} -  m_B(z_i) \}  \Sigma^{-1}_{\text{sne},ij}  \{ m_{B,j} - m_B(z_j) \}  \nonumber \\
&= \{ y_i -  M_B \}  \Sigma^{-1}_{\text{sne},ij}  \{ y_j - M_B \} \,, \\
y_i &=  m_{B,i} -  \mu (z_i) \,,
\end{align}
where the apparent magnitudes $m_{B,i}$, redshifts $z_i$ and covariance matrix $\Sigma_{\rm sne}$ are from the Pantheon catalog (considering both statistical and systematic errors).

In the standard analysis one adopts an improper prior on $M_B$ and integrate over the latter:
\begin{align} \label{marglike}
\mathcal{L}_{\text{sne}}^{\rm marg} & \propto \int_{-\infty}^{+\infty} \d M_B \exp \left [- \frac{1}{2}\chi_{\text{sne}}^{2} \right ] \\
& =  \int_{-\infty}^{+\infty} \d M_B \exp \left [- \frac{1}{2} (  S_2- 2 M_B S_1 +M_B^2 S_0  )  \right ] \nonumber \\
& \propto \exp \left [- \frac{1}{2} \left(  S_{2}-\frac{S_{1}^{2}}{S_{0}}  \right)  \right ]  \,, \nonumber
\end{align}
where inconsequential cosmology-independent factors have been neglected and we defined the auxiliary quantities:
% ----------------------------------------------------------
\begin{align}
S_{0} &=  V_{\text{1}} \cdot \Sigma_{\text{sne}}^{-1} \cdot V_{\text{1}}^{T} \,, \nonumber \\
S_{1} &=  y \cdot \Sigma_{\text{sne}}^{-1} \cdot V_{\text{1}}^{T} \,, \nonumber \\
S_{2} &=  y\cdot \Sigma_{\text{sne}}^{-1} \cdot y^{T} \,,
\end{align}
% ----------------------------------------------------------
where $V_1$ is a vector of unitary elements.
Equivalently, one can use the following new $\chi^2$ function instead of $\chi^2_{\rm sne}$:
\begin{equation} 
\chi_{\text{sne, marg}}^{2}=S_{2}-\frac{S_{1}^{2}}{S_{0}}  \,, \label{chi2SNeMargH}
\end{equation}
which does not depend on $H_0$.

%%%%%%%%%%%%%%%%%%%%%%%%%%%%%%%%%%%%
\subsubsection{Prior on $M_B$} \label{foforma}
%%%%%%%%%%%%%%%%%%%%%%%%%%%%%%%%%%%%

In the case of the $\chi^2_{{\rm tot}, M_B}$ of equation~\eqref{chi2totM}, $M_B$ enters the supernova likelihood and the $M_B$ likelihood, which have to be integrated over at the same time:
\begin{align} \label{marglike}
&\mathcal{L}_{\text{sne+loc}}^{\rm marg}  \propto \int_{-\infty}^{+\infty} \d M_B \exp \left [- \frac{1}{2}(\chi_{\text{sne}}^{2} +\chi_{M_B}^{2} ) \right ]  \\
&= \exp \left [- \frac{1}{2} \left(  S_{2}-\frac{S_{1}^{2}}{S_{0}}  \right)  \right ]  \times \nonumber\\
& \int_{-\infty}^{+\infty} \d  M_B \exp \left [- \frac{1}{2}\frac{\left(M_B - \frac{S_1}{S_0}\right)^2}{S_0^{-1}}  -\frac{1}{2}\frac{\left (M_B-M_B^{\rm R21} \right )^2}{\sigma_{M_B}^2} \right] \nonumber \\
&= \exp \left [- \frac{1}{2} \left(  S_{2}-\frac{S_{1}^{2}}{S_{0}} + \frac{\left(S_1/S_0 - M_B^{\rm R21}\right)^2}{S_0^{-1} + \sigma_{M_B}^2} \right)  \right ] \,, \nonumber
\end{align}
where again inconsequential cosmology-independent factors have been neglected.

Equivalently, one can use the following new $\chi^2$ function instead of $\chi^2_{\rm sne}+ \chi^2_{M_B}$:
%%
%\begin{equation} 
%\chi_{\text{sne+loc, marg}}^{2}=S_{2}-\frac{S_{1}^{2}}{S_{0}} + \frac{\left(S_1/S_0 - M_B^{\rm R21}\right)^2}{S_0^{-1} + \sigma_{M_B}^2}   \,, \label{chi2SNeMargM}
%\end{equation}
%%
%
\begin{align} 
\chi_{\text{sne+loc, marg}}^{2}=\chi_{\text{sne, marg}}^{2}+\chi_{\text{loc}}^{2} \,, \label{chi2SNeMargM}
\end{align}
where
\begin{align}
\chi_{\text{loc}}^{2} = \frac{\left(S_1/S_0 - M_B^{\rm R21}\right)^2}{S_0^{-1} + \sigma_{M_B}^2}   \,.
\end{align}
Note that $\chi_{\text{loc}}^{2}$ does depend on $H_0$.
In particular, it is interesting to consider the case of a diagonal covariance matrix $\Sigma_{\rm sne}= \sigma^2 \mathbb{1}$, where $\mathbb{1}$ is the $n\times n$ identity matrix.
In this case one has:
\begin{align}
\chi_{\text{loc}}^{2} = \frac{\left[ 5 \log_{10}\! H_0 - M_B^{\rm R21} + \frac{1}{n}\sum_i \left( m_{B,i}-5 \log_{10}\!\! \frac{H_0 d_{L, i}}{10\text{pc}} \right)  \right]^2}{\sigma^2/n + \sigma_{M_B}^2}  .
\end{align}
The term within round brackets does not depend on $H_0$ but is a cosmology-dependent intercept that affects the determination of $H_0$ via the local calibration $M_B^{\rm R21}$. Finally, the error is just the sum in quadrature of the calibration and intercept errors.

%%%%%%%%%%%%%%%%%%%%%%%%%%%%%%%%%%%%
%%%%%%%%%%%%%%%%%%%%%%%%%%%%%%%%%%%%
\section{Results}\label{sec:results}
%%%%%%%%%%%%%%%%%%%%%%%%%%%%%%%%%%%%
%%%%%%%%%%%%%%%%%%%%%%%%%%%%%%%%%%%%

%%%%%%%%%%%%%%%%%%%%%%%%%%%%%%%%%%%%%%%%%%%%%%%%%%%%%%    
%\begin{figure*}
%\centering
%\hspace{.2cm}
%\includegraphics[width=0.46 \textwidth]{H0-hsCDM-astro_1}
%\hspace{.9cm}
%\includegraphics[width=0.46 \textwidth]{H0-hsCDM-astro_0}
%\includegraphics[width=0.48 \textwidth]{MB-hsCDM-astro_1}
%\hspace{.5cm}
%\includegraphics[width=0.48 \textwidth]{MB-hsCDM-astro_0}
%\caption{Marginalized constraints %on $\{H_0, \Omega_{M0}, w_x, z_t, M_B, \Omega_{B0}, n_s   \}$
%for hockey-stick dark energy ($hs$CDM) from CMB, BAO, SNe and local observations. The two sets of contours show the analysis that adopts the prior on $M_B$ of equation~\eqref{chi2M} and the one that adopts the prior on $H_0$ of equation~\eqref{chi2H}. As explained in the text, the latter analysis  both biases model selection and distorts the posterior.}
%\label{hs-z}
%\end{figure*}
%%%%%%%%%%%%%%%%%%%%%%%%%%%%%%%%%%%%%%%%%%%%%%%%%%%%%%

%%%%%%%%%%%%%%%%%%%%%%%%%%%%%%%%%%%%%%%%%%%%%%%%%%%%%    
\begin{figure}
\centering
\hspace{.23cm}
\includegraphics[width=.95 \columnwidth]{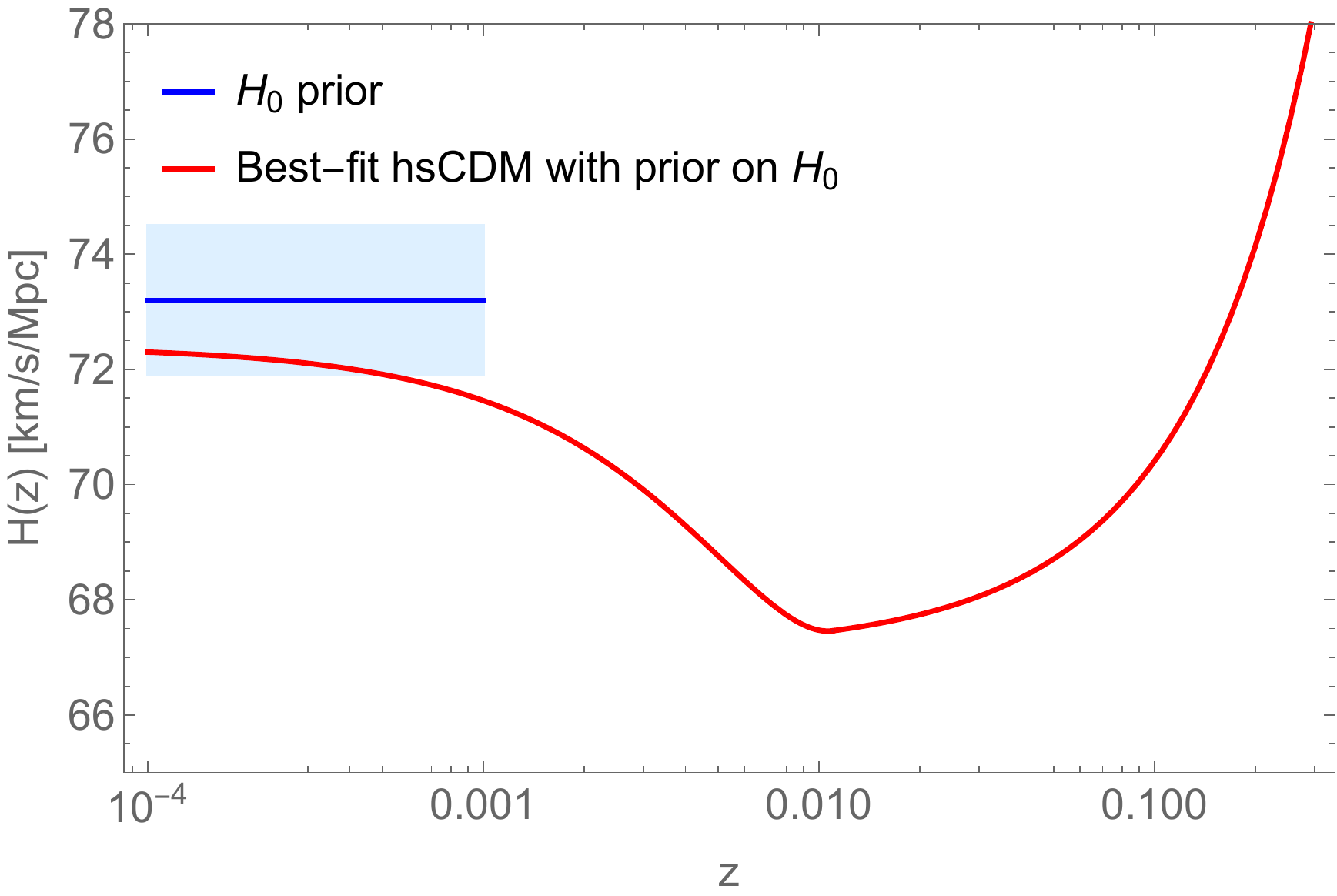}
\includegraphics[width=\columnwidth]{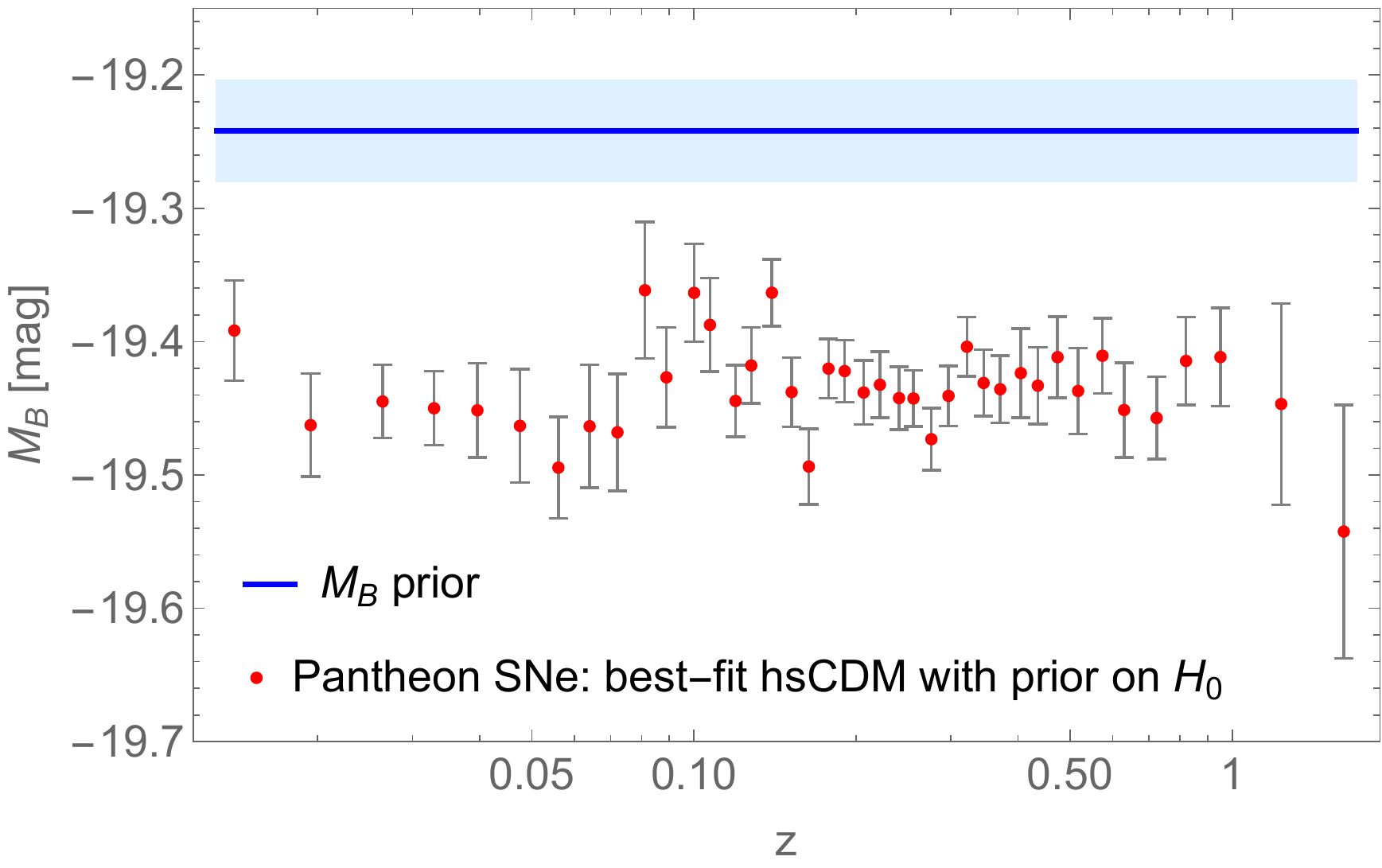}
\caption{
Hubble rate and the inferred absolute magnitudes  $M_{B,i}= m_{B,i}-\mu(z_i) $ for the best-fit $hs$CDM model when using the prior on $H_0$ (Table~\ref{supertab}, top).
For clarity we are showing the binned version of the Pantheon catalog (the statistical analysis uses the full sample).
One can see that, even though the best-fit $H_0$ agrees well with the $H_0$ prior (top), the inferred $M_{B,i}$ do not agree with the local prior on $M_B$ (bottom).}
\label{hs-z}
\end{figure}
%%%%%%%%%%%%%%%%%%%%%%%%%%%%%%%%%%%%%%%%%%%%%%%%%%%%%

%%%%%%%%%%%%%%%%%%%%%%%%%%%%%%%%%%%%%%%%%%%%%%%%%%%%%
\begin{table*}
\begin{center}
\setlength{\tabcolsep}{4pt}
\renewcommand{\arraystretch}{1.5}
\begin{tabular}{>{\centering\arraybackslash}p{1.8cm}|cccc|cc|>{\centering\arraybackslash}p{5.7cm}|>{\centering\arraybackslash}p{1.5cm}>{\centering\arraybackslash}p{1.5cm}}
\hline
\hline
Analysis with  prior on $H_0$ & $\hat \chi^2_{\rm cmb }$& $\hat \chi^2_{\rm bao }$& $\hat \chi^2_{\rm sne }$& $\hat \chi^2_{\rm H_0 }$ & $\hat \chi^2_{\rm tot }$ & $\Delta \hat  \chi^2$ & best-fit vector $\{H_0, \Omega_{M0}, w_x, z_t, M_B, \Omega_{B0}, n_s   \}$& distance from $H^{\rm R21}_0$ & distance from $M^{\rm R21}_B$\\ \hline
$w$CDM & 2.9 & 5.1 & 1030.0 & 7.8 & 1045.8 & 0  & \{69.6, 0.29, -1.08, ------, -19.39, 0.046, 0.97\} & 2.8 & 3.8 \\
$hs$CDM & 1.3 & 5.9 & 1027.7 & 0.3 & 1035.1 &  -10.7 & \{72.5, 0.26, -14.4, 0.010, -19.42, 0.043, 0.97\} & 0.5 & 4.9 \\
\hline
\hline
Analysis with  prior on $M_B$ & $\hat \chi^2_{\rm cmb }$& $\hat \chi^2_{\rm bao }$& $\hat \chi^2_{\rm sne }$& $\hat \chi^2_{\rm M_B }$ & $\hat \chi^2_{\rm tot }$ & $\Delta \hat \chi^2$ & best-fit vector $\{H_0, \Omega_{M0}, w_x, z_t, M_B, \Omega_{B0}, n_s   \}$& distance from $H^{\rm R21}_0$ & distance from $M^{\rm R21}_B$\\ \hline
$w$CDM  & 2.8 & 5.2 & 1029.3 & 14.6 & 1051.9 & 0 & \{69.4, 0.29, -1.07, \;-----, -19.39, 0.047, 0.97\} & 2.9 & 3.8 \\
$hs$CDM  & 1.8 & 7.1 & 1027.1 & 19.4 & 1055.4 & 3.5 & \{69.3, 0.29, -1.73, 0.055, -19.41, 0.047, 0.97\} & 3.0 & 4.4 \\
\hline
\hline
\end{tabular}
\caption{
Comparison between the best fits relative to the analyses of equations~\eqref{chi2totH} (top) and~\eqref{chi2totM} (bottom).
A hat denotes the minimum $\chi^2$. The $\Delta \hat  \chi^2$ values are computed with respect to $w$CDM.
The last two columns give the $\sigma$-distance $(H^{\rm R21}_0-H_0^{\rm bf})/\sigma_{H_0^{\rm R21}}$ and $(M^{\rm R21}_B-M_B^{\rm bf})/\sigma_{M_B^{\rm R21}}$ from the values given in equation~\eqref{R21} and Table~\eqref{calib} (Pantheon), respectively.
}
\label{supertab}
\end{center}
\end{table*}
%%%%%%%%%%%%%%%%%%%%%%%%%%%%%%%%%%%%%%%%%%%%%%%%%%%%%

Table~\ref{supertab} shows a comparison between the best fits relative to the analyses of equations~\eqref{chi2totH} and~\eqref{chi2totM}.
Our results are obtained using the numerical codes \texttt{CLASS} \citep{Blas:2011rf}, \texttt{MontePython} \citep{Audren:2012wb} and \texttt{getdist} \citep{Lewis:2019xzd}.

When using the prior on $H_0$ (Table~\ref{supertab}, top), the $hs$CDM model, with an extremely phantom $w_x\simeq -14$, features a significantly lower minimum $\chi^2$ as compared to the $w$CDM model. In particular, the disagreement with respect to the SH0ES determination of equation~\eqref{R21} is completely resolved. The phantom transition seems to have explained away the $H_0$ crisis.
However, the best-fit $M_B$ is 5$\sigma$ away from the prior on $M_B$ from Table~\eqref{calib}, and this information is not included in the total $\chi^2$. This biases both model selection and the best-fit model.
To better illustrate this point, we show in Figure~\ref{hs-z} the Hubble rate and the inferred absolute magnitudes  $M_{B,i}= m_{B,i}-\mu(z_i)$ for the best-fit $hs$CDM model.
Even though the best-fit $H_0$ agrees well with the $H_0$ prior (Figure~\ref{hs-z}, top), the inferred $M_{B,i}$ do not agree with the local prior on $M_B$ throughout the full redshift range (Figure~\ref{hs-z}, bottom).

When, instead, the $\chi^2_{M_B}$ of equation~\eqref{chi2M} is adopted (Table~\ref{supertab}, bottom), the $hs$CDM model features the same best-fit $H_0$ of the $w$CDM model, both 3$\sigma$ away from the SH0ES determination of equation~\eqref{R21}.
Moreover, the $hs$CDM has a worse overall fit to the data as compared to $w$CDM.
In other words, hockey-stick dark energy neither solves the $H_0$ crisis nor manifests any statistical advantage with respect to $w$CDM.

%%%%%%%%%%%%%%%%%%%%%%%%%%%%%%%%%%%%%%%%%%%%%%%%%%%%%    
\begin{figure}
\centering
\includegraphics[width=0.5\textwidth]{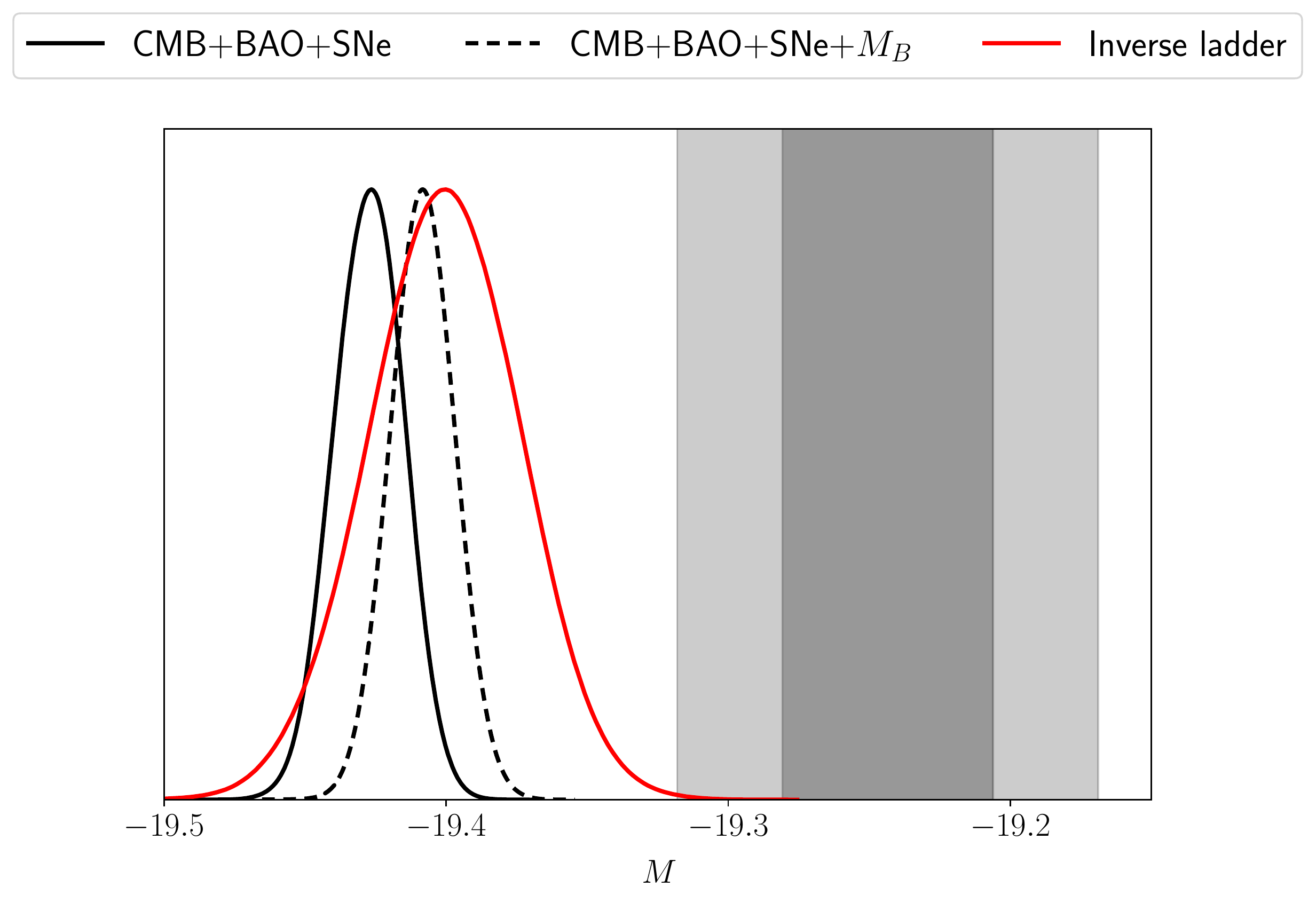}
\caption{The posteriors on $M_B$ from the analyses of the $hs$CDM model (black line) and the inverse-distance ladder (red line) are in disagreement with the local prior from Table~\ref{calib} (grey contour).
This robustly shows that the SN calibration produced by CMB and BAO is in tension with the local astrophysical calibration. This remains true  if a local prior on $M_B$ is used in the analysis of the $hs$CDM model (black dashed line): the difficulty in matching the local calibration is the source of the $H_0$ crisis.}
\label{3M}
\end{figure}
%%%%%%%%%%%%%%%%%%%%%%%%%%%%%%%%%%%%%%%%%%%%%%%%%%%%%

From these results it is clear what is the source of the Hubble crisis.
CMB and BAO constrain tightly the luminosity distance-redshift relation and so the distance modulus $\mu(z)$. The Pantheon dataset constrains the supernova apparent magnitudes $m_B$. Consequently, CMB, BAO and SNe produce a calibration on $M_B$ which happens to be in strong disagreement with the local astrophysical calibration via Cepheids (see Figure~\ref{hs-z} and Table~\ref{supertab}).
This disagreement was highlighted by \citet[][Figure 5]{Camarena:2019rmj} where the inverse-distance ladder technique was used to propagate the CMB constraint on $r_d$ to $M_B$ in a parametric-free way.
Figure~\ref{3M} shows how the constraint on $M_B$ from the inverse-distance ladder analysis agrees with the one relative to the $hs$CDM model.

%%%%%%%%%%%%%%%%%%%%%%%%%%%%%%%%%%%%%%%%%%%%%%%%%%%%%    
\begin{figure*}
\centering
\includegraphics[width=\textwidth]{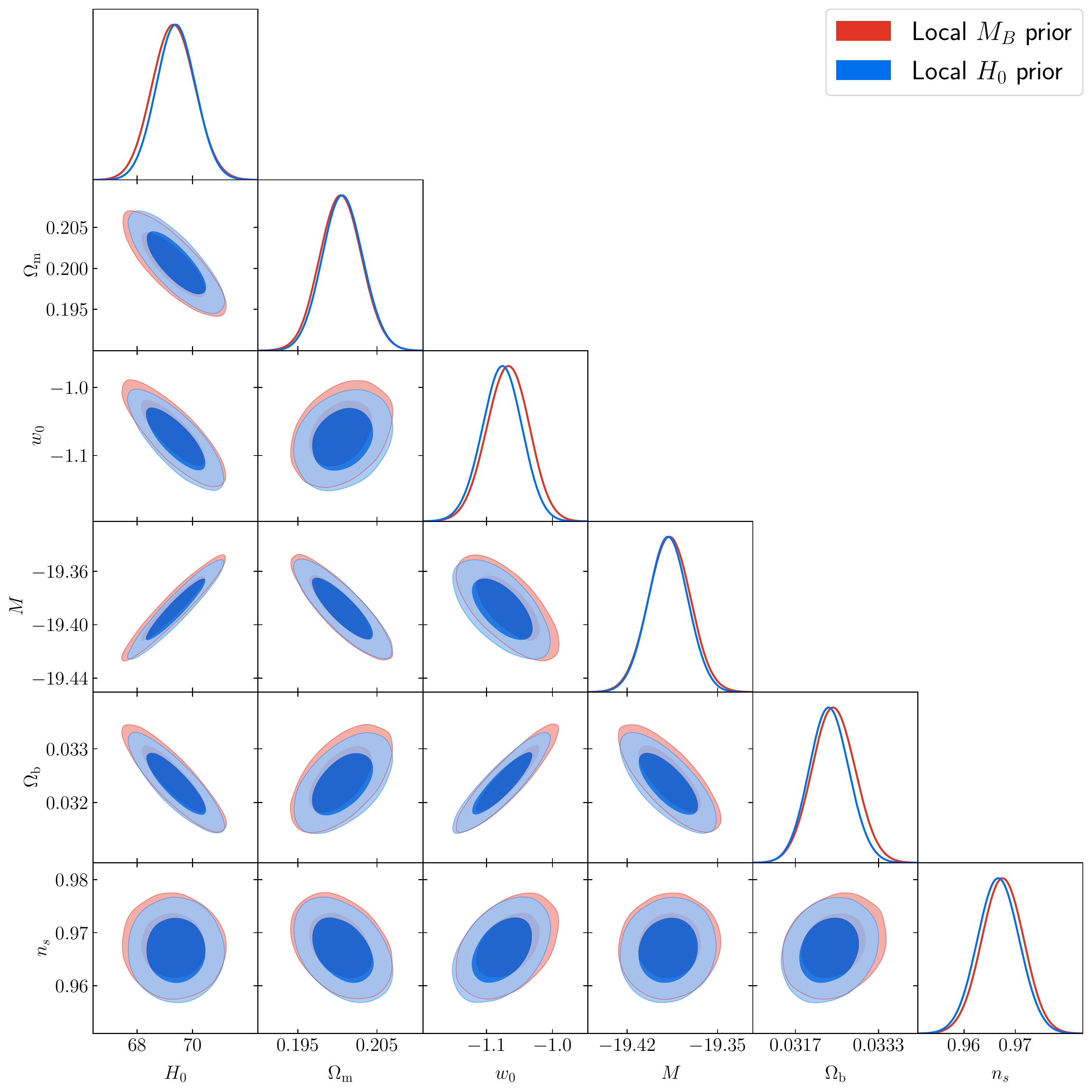}
\caption{Marginalized constraints %on $\{H_0, \Omega_{M0}, w_x, z_t, M_B, \Omega_{B0}, n_s   \}$
for the $w$CDM model from CMB, BAO, SNe and local observations. The two sets of contours show the analysis that adopts the prior on $M_B$ of equation~\eqref{chi2M} and the one that adopts the prior on $H_0$ of equation~\eqref{chi2H}.}
\label{w-bayes}
\end{figure*}
%%%%%%%%%%%%%%%%%%%%%%%%%%%%%%%%%%%%%%%%%%%%%%%%%%%%%

%%%%%%%%%%%%%%%%%%%%%%%%%%%%%%%%%%%%%%%%%%%%%%%%%%%%%    
\begin{figure*}
\centering
\includegraphics[width=\textwidth]{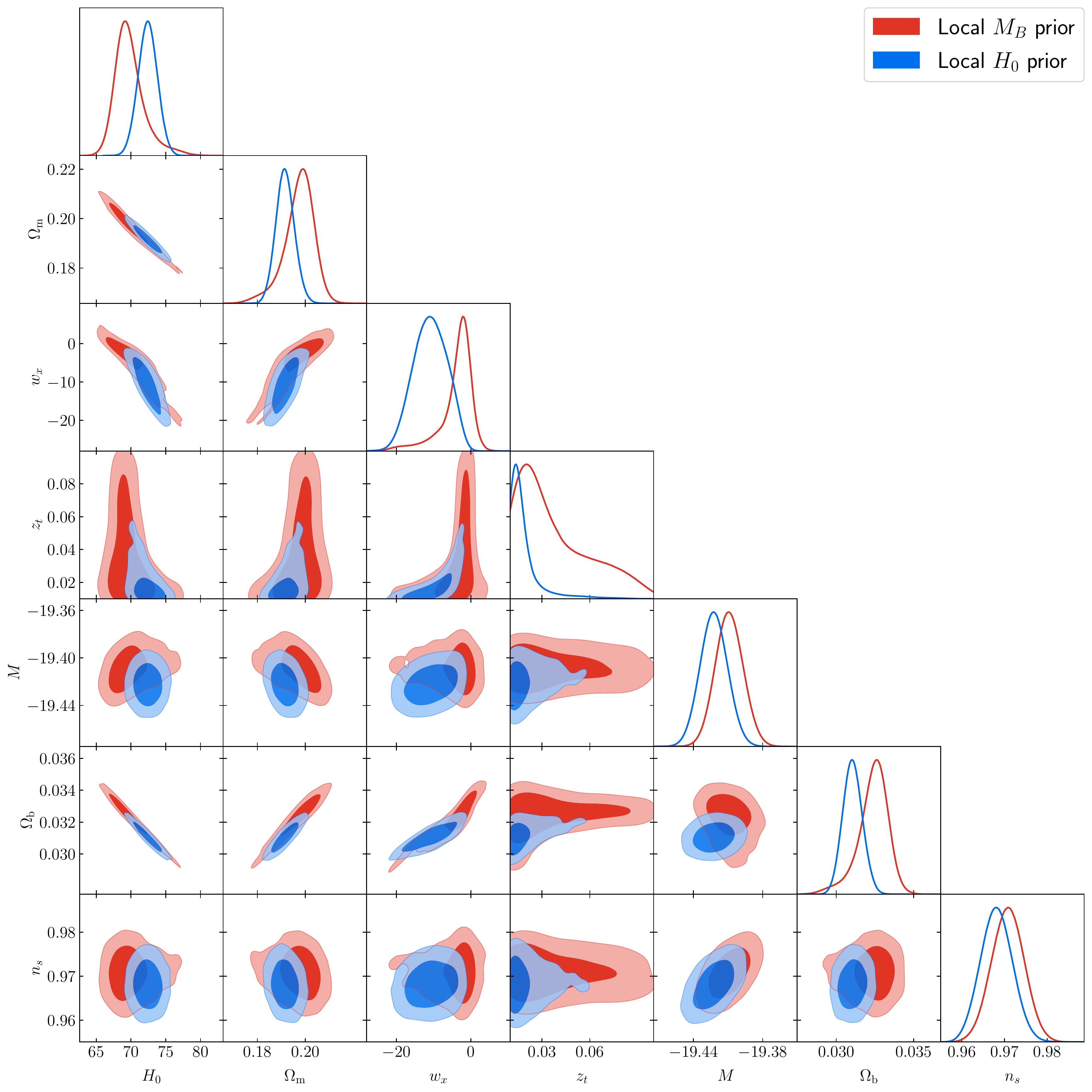}
\caption{Marginalized constraints %on $\{H_0, \Omega_{M0}, w_x, z_t, M_B, \Omega_{B0}, n_s   \}$
for hockey-stick dark energy ($hs$CDM) from CMB, BAO, SNe and local observations. The two sets of contours show the analysis that adopts the prior on $M_B$ of equation~\eqref{chi2M} and the one that adopts the prior on $H_0$ of equation~\eqref{chi2H}. As explained in the text, the latter analysis  both biases model selection and distorts the posterior.}
\label{hs-bayes}
\end{figure*}
%%%%%%%%%%%%%%%%%%%%%%%%%%%%%%%%%%%%%%%%%%%%%%%%%%%%%

Finally, Figures~\ref{w-bayes} and~\ref{hs-bayes} show how Bayesian inference changes when one adopts the prior on $M_B$ instead of the prior on $H_0$.
The impact on the analysis relative to $w$CDM is minimal, suggesting the validity of previous analyses of the $w$CDM model that adopted the prior on $H_0$.
On the other hand, the constraints relative to the $hs$CDM model change significantly. The impact on $w_x$ is particular strong: in the case of the analysis with the prior on $H_0$ much more phantom values of $w_x$ are allowed as compared with the analysis with the prior on $M_B$.
Also note that the analysis with $M_B$ includes $z_t = 0.1$ at $2\sigma$ level while the analysis with $H_0$ constrains $z_t < 0.06$ at $2\sigma$ level, showing a preference for a low-redshift transition.
In other words, in the case of models with a low-redshift transition, the use of the prior on $H_0$ both biases model selection and distorts the posterior.
In the analysis we adopted the flat prior $0.01\le  z_t \le 0.1$ \citep{Benevento:2020fev,Alestas:2020zol}: a transition at redshifts lower than 0.01 would not affect the determination of $H_0$ and a transition at redshifts higher than 0.1 would not solve the $H_0$ crisis, as also shown by  Figure~\ref{hs-bayes} (blue curve).

%%%%%%%%%%%%%%%%%%%%%%%%%%%%%%%%%%%%
%%%%%%%%%%%%%%%%%%%%%%%%%%%%%%%%%%%%
\section{Conclusions} \label{sec:conclu}
%%%%%%%%%%%%%%%%%%%%%%%%%%%%%%%%%%%%
%%%%%%%%%%%%%%%%%%%%%%%%%%%%%%%%%%%%

In this paper we clearly show that a sudden phantom transition at very-low redshift cannot solve the >4$\sigma$ disagreement between the local and high-redshift determinations of the Hubble constant.
This point has been previously made by \citet{Benevento:2020fev} in the contest of a sudden low-redshift discontinuity in the expansion rate, and by \citet{Lemos:2018smw} who showed through an $H(z)$ reconstruction that SN, BAO and $r_d$ constraints do not allow for a higher expansion rate at low redshifts \citep[see also the recent analysis by][]{Efstathiou:2021ocp}.\footnote{Similar conclusions can be derived from the non-parameter inverse distance ladder analysis of \citet{Camarena:2019rmj}, which shows that the calibration given by CMB and BAO to SN does not agree with the one by Cepheid distances, see Fig.~\ref{3M}.}

Here, we single out the reason of this failure in solving the $H_0$ crisis: the supernova absolute magnitude $M_B$ that is used to derive the local $H_0$ constraint is not compatible with the $M_B$ that is necessary to fit supernova, BAO and CMB data, see Figures~\ref{hs-z} and~\ref{3M}.
Statistically, this incompatibility is taken into account in the analysis if one adopts the supernova calibration prior on~$M_B$ instead of the prior on~$H_0$.

For completeness, we wish to summarize the three reasons why one should use the $\chi^2_{M_B}$ of equation~\eqref{chi2M} instead of the $\chi^2_{H_0}$ of equation~\eqref{chi2H}:%
\footnote{Some of these points were previously raised by \citet{Camarena:2019moy,Benevento:2020fev}.}
\begin{enumerate}

\item The use of $\chi^2_{M_B}$ avoids potential double counting low-redshift supernovae: 
for example, there are 175 supernovae in common between the Supercal and Pantheon datasets in the range $0.023\le z\le 0.15$, and, in the standard analysis, these supernovae are used twice: once for the $H_0$ determination and once when constraining the cosmological parameters.
This induces a covariance between $H_0$ and the other parameters which could bias  cosmological inference.

\item  The supernova calibration prior on $M_B$ is an astrophysical and local measurement.
The determination of $H_0$ is instead based on a cosmographic analysis and it depends on a) its validity and b) its priors (SH0ES adopts $q_0=-0.55$). While one can relax b) and obtain a joint $H_0$-$q_0$ prior (see Table~\ref{CM20}), the cosmographic analysis  may fail for models with sudden transitions such as $hs$CDM.
As $\chi^2_{M_B}$ is not based on a cosmographic analysis, it does not suffer from these issues.

\item Most importantly, the use of $\chi^2_{M_B}$ guarantees that one includes in the analysis the fact that $M_B$ is constrained by the calibration prior of Table~\eqref{calib}.

\end{enumerate}

While, as shown by Table~\ref{supertab} and Figures~\ref{w-bayes} and~\ref{hs-bayes}, the conclusions for $w$CDM are not changed when $\chi^2_{M_B}$ is adopted, the use of $\chi^2_{M_B}$ becomes compelling when more exotic models are investigated; the use of $\chi^2_{H_0}$ can indeed lead to incorrect conclusions.
Given that using $\chi^2_{M_B}$ does not add any statistical complexity to the analysis (see equation~\eqref{chi2SNeMargM}), we encourage the community to adopt this prior in all the analyses.
However, one should bear in mind that one must adopt the prior on $M_B$ that corresponds to the supernova dataset that one wishes to adopt in their statistical analysis.
These can be obtained using the code made available at \href{https://github.com/valerio-marra/CalPriorSNIa}{github.com/valerio-marra/CalPriorSNIa}, where we will keep an updated list of the $M_B$ priors that correspond to the latest supernova catalogs.

%%%%%%%%%%%%%%%%%%%%%%%%%%%%%%%%%%%%
%%%%%%%%%%%%%%%%%%%%%%%%%%%%%%%%%%%%
\section*{Acknowledgements}
It is a pleasure to thank Leandros Perivolaropoulos, Adam Riess, Sunny Vagnozzi and Adrià Gómez-Valent for useful comments and discussions.
DC thanks CAPES for financial support.
VM thanks CNPq and FAPES for partial financial support.
This project has received funding from the European Union’s Horizon 2020 research and innovation programme under the Marie Skłodowska-Curie grant agreement No 888258.
This work also made use of the Virgo Cluster at Cosmo-ufes/UFES, which is funded by FAPES and administrated by Renan Alves de Oliveira.

%%%%%%%%%%%%%%%%%%%%%%%%%%%%%%%%%%%%
%%%%%%%%%%%%%%%%%%%%%%%%%%%%%%%%%%%%

\section*{Data availability}

The data underlying this article will be shared on reasonable request to the corresponding author.

%%%%%%%%%%%%%%%%%%%%%%%%%%%%%%%%%%%%
%%%%%%%%%%%%% REFERENCES %%%%%%%%%%%%%%%
\bibliographystyle{mnrasArxiv}
\bibliography{biblio}

\begin{thebibliography}{}
\makeatletter
\relax
\def\mn@urlcharsother{\let\do\@makeother \do\$\do\&\do\#\do\^\do\_\do\%\do\~}
\def\mn@doi{\begingroup\mn@urlcharsother \@ifnextchar [ {\mn@doi@}
  {\mn@doi@[]}}
\def\mn@doi@[#1]#2{\def\@tempa{#1}\ifx\@tempa\@empty \href
  {http://dx.doi.org/#2} {doi:#2}\else \href {http://dx.doi.org/#2} {#1}\fi
  \endgroup}
\def\mn@eprint#1#2{\mn@eprint@#1:#2::\@nil}
\def\mn@eprint@arXiv#1{\href {http://arxiv.org/abs/#1} {{\tt arXiv:#1}}}
\def\mn@eprint@dblp#1{\href {http://dblp.uni-trier.de/rec/bibtex/#1.xml}
  {dblp:#1}}
\def\mn@eprint@#1:#2:#3:#4\@nil{\def\@tempa {#1}\def\@tempb {#2}\def\@tempc
  {#3}\ifx \@tempc \@empty \let \@tempc \@tempb \let \@tempb \@tempa \fi \ifx
  \@tempb \@empty \def\@tempb {arXiv}\fi \@ifundefined
  {mn@eprint@\@tempb}{\@tempb:\@tempc}{\expandafter \expandafter \csname
  mn@eprint@\@tempb\endcsname \expandafter{\@tempc}}}

\bibitem[\protect\citeauthoryear{Aghanim et~al.}{Aghanim
  et~al.}{2020}]{Aghanim:2018eyx}
Aghanim N.,  et~al., 2020, \mn@doi [Astron. Astrophys.]
  {10.1051/0004-6361/201833910}, 641, A6,
  [\href{https://arxiv.org/abs/1807.06209}{1807.06209}].

\bibitem[\protect\citeauthoryear{Alam et~al.}{Alam et~al.}{2017}]{Alam:2016hwk}
Alam S.,  et~al., 2017, \mn@doi [Mon. Not. Roy. Astron. Soc.]
  {10.1093/mnras/stx721}, 470, 2617,
  [\href{https://arxiv.org/abs/1607.03155}{1607.03155}].

\bibitem[\protect\citeauthoryear{Alestas, Kazantzidis  \&
  Perivolaropoulos}{Alestas et~al.}{2020}]{Alestas:2020zol}
Alestas G.,  Kazantzidis L.,   Perivolaropoulos L.,  2020,
  [\href{https://arxiv.org/abs/2012.13932}{2012.13932}].

\bibitem[\protect\citeauthoryear{Audren, Lesgourgues, Benabed  \&
  Prunet}{Audren et~al.}{2013}]{Audren:2012wb}
Audren B.,  Lesgourgues J.,  Benabed K.,   Prunet S.,  2013, \mn@doi [JCAP]
  {10.1088/1475-7516/2013/02/001}, 1302, 001,
  [\href{https://arxiv.org/abs/1210.7183}{1210.7183}].

\bibitem[\protect\citeauthoryear{Benevento, Hu  \& Raveri}{Benevento
  et~al.}{2020}]{Benevento:2020fev}
Benevento G.,  Hu W.,   Raveri M.,  2020, \mn@doi [Phys. Rev. D]
  {10.1103/PhysRevD.101.103517}, 101, 103517,
  [\href{https://arxiv.org/abs/2002.11707}{2002.11707}].

\bibitem[\protect\citeauthoryear{Beutler et~al.,}{Beutler
  et~al.}{2011}]{Beutler:2011hx}
Beutler F.,  et~al., 2011, \mn@doi [Mon. Not. Roy. Astron. Soc.]
  {10.1111/j.1365-2966.2011.19250.x}, 416, 3017,
  [\href{https://arxiv.org/abs/1106.3366}{1106.3366}].

\bibitem[\protect\citeauthoryear{Blas, Lesgourgues  \& Tram}{Blas
  et~al.}{2011}]{Blas:2011rf}
Blas D.,  Lesgourgues J.,   Tram T.,  2011, \mn@doi [JCAP]
  {10.1088/1475-7516/2011/07/034}, 1107, 034,
  [\href{https://arxiv.org/abs/1104.2933}{1104.2933}].

\bibitem[\protect\citeauthoryear{Brout \& Scolnic}{Brout \&
  Scolnic}{2021}]{Brout:2020msh}
Brout D.,  Scolnic D.,  2021, \mn@doi [Astrophys. J.]
  {10.3847/1538-4357/abd69b}, 909, 26,
  [\href{https://arxiv.org/abs/2004.10206}{2004.10206}].

\bibitem[\protect\citeauthoryear{Brout et~al.}{Brout
  et~al.}{2019}]{Brout:2018jch}
Brout D.,  et~al., 2019, \mn@doi [Astrophys. J.] {10.3847/1538-4357/ab08a0},
  874, 150, [\href{https://arxiv.org/abs/1811.02377}{1811.02377}].

\bibitem[\protect\citeauthoryear{Camarena \& Marra}{Camarena \&
  Marra}{2020a}]{Camarena:2019moy}
Camarena D.,  Marra V.,  2020a, \mn@doi [Phys. Rev. Res.]
  {10.1103/PhysRevResearch.2.013028}, 2, 013028,
  [\href{https://arxiv.org/abs/1906.11814}{1906.11814}].

\bibitem[\protect\citeauthoryear{Camarena \& Marra}{Camarena \&
  Marra}{2020b}]{Camarena:2019rmj}
Camarena D.,  Marra V.,  2020b, \mn@doi [Mon. Not. Roy. Astron. Soc.]
  {10.1093/mnras/staa770}, 495, 2630,
  [\href{https://arxiv.org/abs/1910.14125}{1910.14125}].

\bibitem[\protect\citeauthoryear{Chen, Huang  \& Wang}{Chen
  et~al.}{2019}]{Chen:2018dbv}
Chen L.,  Huang Q.-G.,   Wang K.,  2019, \mn@doi [JCAP]
  {10.1088/1475-7516/2019/02/028}, 02, 028,
  [\href{https://arxiv.org/abs/1808.05724}{1808.05724}].

\bibitem[\protect\citeauthoryear{Dhawan, Brout, Scolnic, Goobar, Riess  \&
  Miranda}{Dhawan et~al.}{2020}]{Dhawan:2020xmp}
Dhawan S.,  Brout D.,  Scolnic D.,  Goobar A.,  Riess A.,   Miranda V.,  2020,
  \mn@doi [Astrophys. J.] {10.3847/1538-4357/ab7fb0}, 894, 54,
  [\href{https://arxiv.org/abs/2001.09260}{2001.09260}].

\bibitem[\protect\citeauthoryear{Di~Valentino et~al.,}{Di~Valentino
  et~al.}{2021}]{DiValentino:2021izs}
Di~Valentino E.,  et~al., 2021,
  [\href{https://arxiv.org/abs/2103.01183}{2103.01183}].

\bibitem[\protect\citeauthoryear{Efstathiou}{Efstathiou}{2021}]{Efstathiou:2021ocp}
Efstathiou G.,  2021, [\href{https://arxiv.org/abs/2103.08723}{2103.08723}].

\bibitem[\protect\citeauthoryear{Foreman-Mackey, Hogg, Lang  \&
  Goodman}{Foreman-Mackey et~al.}{2013}]{ForemanMackey:2012ig}
Foreman-Mackey D.,  Hogg D.~W.,  Lang D.,   Goodman J.,  2013, \mn@doi [Publ.
  Astron. Soc. Pac.] {10.1086/670067}, 125, 306,
  [\href{https://arxiv.org/abs/1202.3665}{1202.3665}].

\bibitem[\protect\citeauthoryear{Goliath, Amanullah, Astier, Goobar  \&
  Pain}{Goliath et~al.}{2001}]{Goliath:2001af}
Goliath M.,  Amanullah R.,  Astier P.,  Goobar A.,   Pain R.,  2001, \mn@doi
  [Astron. Astrophys.] {10.1051/0004-6361:20011398}, 380, 6,
  [\href{https://arxiv.org/abs/astro-ph/0104009}{astro-ph/0104009}].

\bibitem[\protect\citeauthoryear{Guy et~al.}{Guy et~al.}{2007}]{Guy:2007dv}
Guy J.,  et~al., 2007, \mn@doi [Astron. Astrophys.]
  {10.1051/0004-6361:20066930}, 466, 11,
  [\href{https://arxiv.org/abs/astro-ph/0701828}{astro-ph/0701828}].

\bibitem[\protect\citeauthoryear{Huang}{Huang}{2020}]{Huang:2020mub}
Huang Z.,  2020, \mn@doi [Astrophys. J. Lett.] {10.3847/2041-8213/ab8011}, 892,
  L28, [\href{https://arxiv.org/abs/2001.06926}{2001.06926}].

\bibitem[\protect\citeauthoryear{Kang, Lee, Kim, Chung  \& Ree}{Kang
  et~al.}{2020}]{Kang:2019azh}
Kang Y.,  Lee Y.-W.,  Kim Y.-L.,  Chung C.,   Ree C.~H.,  2020, \mn@doi
  [Astrophys. J.] {10.3847/1538-4357/ab5afc}, 889, 8,
  [\href{https://arxiv.org/abs/1912.04903}{1912.04903}].

\bibitem[\protect\citeauthoryear{Keeley, Joudaki, Kaplinghat  \& Kirkby}{Keeley
  et~al.}{2019}]{Keeley:2019esp}
Keeley R.~E.,  Joudaki S.,  Kaplinghat M.,   Kirkby D.,  2019, \mn@doi [JCAP]
  {10.1088/1475-7516/2019/12/035}, 12, 035,
  [\href{https://arxiv.org/abs/1905.10198}{1905.10198}].

\bibitem[\protect\citeauthoryear{{Kelly}, {Hicken}, {Burke}, {Mandel}  \&
  {Kirshner}}{{Kelly} et~al.}{2010}]{2010ApJ...715..743K}
{Kelly} P.~L.,  {Hicken} M.,  {Burke} D.~L.,  {Mandel} K.~S.,   {Kirshner}
  R.~P.,  2010, \mn@doi [\apj] {10.1088/0004-637X/715/2/743}, 715, 743,
  [\href{https://arxiv.org/abs/0912.0929}{0912.0929}].

\bibitem[\protect\citeauthoryear{Kim, Kang  \& Lee}{Kim
  et~al.}{2019}]{Kim:2019npy}
Kim Y.-L.,  Kang Y.,   Lee Y.-W.,  2019, \mn@doi [J. Korean Astron. Soc.]
  {10.5303/JKAS.2019.52.5.181}, 52, 181,
  [\href{https://arxiv.org/abs/1908.10375}{1908.10375}].

\bibitem[\protect\citeauthoryear{Knox \& Millea}{Knox \&
  Millea}{2020}]{Knox:2019rjx}
Knox L.,  Millea M.,  2020, \mn@doi [Phys. Rev. D]
  {10.1103/PhysRevD.101.043533}, 101, 043533,
  [\href{https://arxiv.org/abs/1908.03663}{1908.03663}].

\bibitem[\protect\citeauthoryear{Koo, Shafieloo, Keeley  \& L'Huillier}{Koo
  et~al.}{2020}]{Koo:2020ssl}
Koo H.,  Shafieloo A.,  Keeley R.~E.,   L'Huillier B.,  2020, \mn@doi
  [Astrophys. J.] {10.3847/1538-4357/ab9c9a}, 899, 9,
  [\href{https://arxiv.org/abs/2001.10887}{2001.10887}].

\bibitem[\protect\citeauthoryear{{Lampeitl} et~al.,}{{Lampeitl}
  et~al.}{2010}]{2010ApJ...722..566L}
{Lampeitl} H.,  et~al., 2010, \mn@doi [\apj] {10.1088/0004-637X/722/1/566},
  722, 566, [\href{https://arxiv.org/abs/1005.4687}{1005.4687}].

\bibitem[\protect\citeauthoryear{Lemos, Lee, Efstathiou  \& Gratton}{Lemos
  et~al.}{2019}]{Lemos:2018smw}
Lemos P.,  Lee E.,  Efstathiou G.,   Gratton S.,  2019, \mn@doi [Mon. Not. Roy.
  Astron. Soc.] {10.1093/mnras/sty3082}, 483, 4803,
  [\href{https://arxiv.org/abs/1806.06781}{1806.06781}].

\bibitem[\protect\citeauthoryear{Lewis}{Lewis}{2019}]{Lewis:2019xzd}
Lewis A.,  2019, [\href{https://arxiv.org/abs/1910.13970}{1910.13970}].

\bibitem[\protect\citeauthoryear{Mortonson, Hu  \& Huterer}{Mortonson
  et~al.}{2009}]{Mortonson:2009qq}
Mortonson M.~J.,  Hu W.,   Huterer D.,  2009, \mn@doi [Phys. Rev. D]
  {10.1103/PhysRevD.80.067301}, 80, 067301,
  [\href{https://arxiv.org/abs/0908.1408}{0908.1408}].

\bibitem[\protect\citeauthoryear{Riess et~al.}{Riess
  et~al.}{2016}]{Riess:2016jrr}
Riess A.~G.,  et~al., 2016, \mn@doi [Astrophys. J.]
  {10.3847/0004-637X/826/1/56}, 826, 56,
  [\href{https://arxiv.org/abs/1604.01424}{1604.01424}].

\bibitem[\protect\citeauthoryear{Riess, Casertano, Yuan, Bowers, Macri, Zinn
  \& Scolnic}{Riess et~al.}{2021}]{Riess:2020fzl}
Riess A.~G.,  Casertano S.,  Yuan W.,  Bowers J.~B.,  Macri L.,  Zinn J.~C.,
  Scolnic D.,  2021, \mn@doi [Astrophys. J. Lett.] {10.3847/2041-8213/abdbaf},
  908, L6, [\href{https://arxiv.org/abs/2012.08534}{2012.08534}].

\bibitem[\protect\citeauthoryear{Rose et~al.,}{Rose
  et~al.}{2020}]{Rose:2020shp}
Rose B.~M.,  et~al., 2020, \mn@doi [Astrophys. J. Lett.]
  {10.3847/2041-8213/ab94ad}, 896, L4,
  [\href{https://arxiv.org/abs/2002.12382}{2002.12382}].

\bibitem[\protect\citeauthoryear{Ross, Samushia, Howlett, Percival, Burden  \&
  Manera}{Ross et~al.}{2015}]{Ross:2014qpa}
Ross A.~J.,  Samushia L.,  Howlett C.,  Percival W.~J.,  Burden A.,   Manera
  M.,  2015, \mn@doi [Mon. Not. Roy. Astron. Soc.] {10.1093/mnras/stv154}, 449,
  835, [\href{https://arxiv.org/abs/1409.3242}{1409.3242}].

\bibitem[\protect\citeauthoryear{Sapone, Nesseris  \& Bengaly}{Sapone
  et~al.}{2020}]{Sapone:2020wwz}
Sapone D.,  Nesseris S.,   Bengaly C. A.~P.,  2020,
  [\href{https://arxiv.org/abs/2006.05461}{2006.05461}].

\bibitem[\protect\citeauthoryear{Scolnic et~al.}{Scolnic
  et~al.}{2015}]{Scolnic:2015eyc}
Scolnic D.,  et~al., 2015, \mn@doi [Astrophys. J.]
  {10.1088/0004-637X/815/2/117}, 815, 117,
  [\href{https://arxiv.org/abs/1508.05361}{1508.05361}].

\bibitem[\protect\citeauthoryear{Scolnic et~al.}{Scolnic
  et~al.}{2018}]{Scolnic:2017caz}
Scolnic D.~M.,  et~al., 2018, \mn@doi [Astrophys. J.]
  {10.3847/1538-4357/aab9bb}, 859, 101,
  [\href{https://arxiv.org/abs/1710.00845}{1710.00845}].

\bibitem[\protect\citeauthoryear{{Sullivan} et~al.,}{{Sullivan}
  et~al.}{2010}]{2010MNRAS.406..782S}
{Sullivan} M.,  et~al., 2010, \mn@doi [\mnras]
  {10.1111/j.1365-2966.2010.16731.x}, 406, 782,
  [\href{https://arxiv.org/abs/1003.5119}{1003.5119}].

\makeatother
\end{thebibliography}
%%%%%%%%%%%%%%%%%%%%%%%%%%%%%%%%%%%%
%%%%%%%%%%%%%%%%%%%%%%%%%%%%%%%%%%%%

%%%%%%%%%%%%%%%%%%%%%%%%%%%%%%%%%%%%%%%%%%%%%%%%%%

% Don't change these lines
\bsp	% typesetting comment
\label{lastpage}
\end{document}